# Multicolour wavelength-tunable lasing from a single bandgap-graded alloy nanoribbon

Yize Lu, Fuxing Gu, Chao Meng, Huakang Yu, Yaoguang Ma, Wei Fang,[a] and Limin Tong

*State Key Laboratory of Modern Optical Instrumentation, Department of Optical Engineering, Zhejiang University, Hangzhou 310027, China*

[a] Electronic mail: wfang08@zju.edu.cn

Tunable lasing from 578 nm to 640 nm is observed from a single CdSSe bandgap-graded alloy nanoribbon, by selecting the excited spot at room temperature. Though reabsorption is a serious problem to achieve lasing at short wavelength, multiple scatters on the nanoribbon form localized cavities, and thus realize lasing at different wavelengths. By increasing the excitation area, we also observe multicolour lasing from the same nanoribbon simultaneously.

One-dimensional semiconductor micro-nanowire (MNW) is a promising building block for integrated photonic applications with compact size, strong optical confinement, sufficient optical gain, and easy synthesis.[1] Various kinds of devices based on single MNW have been realized, including lasers,[2] light-emitting diodes,[3] active waveguides,[4,5] photodetectors,[6] and photonic biosensors.[7] However, the factor that the majority MNW devices are made of homogeneous materials, thus mono-bandgaps, has limited many applications. In the past decade, the emerging of spatial bandgap-graded alloy semiconductor MNWs with composition-tunable gain has attract intense attentions.[8-13] More recently, the realization of wide-range bandgap tenability in a single $CdS_{1-x}Se_x$ MNW structure has expended the application horizon greatly.[14-16] In such single crystalline structure, the composition along the length of MNW can be continuously varied from $x = 1$ at one end to $x = 0$ at the other end, and consequently the bandgap is tuned from 1.74 eV to 2.44 eV. The broad emission/absorption spectra make ultrafast nanolaser, white light emitting diode, full-spectrum solar cell possible from a single nanowire structure.

In this work, we present a multicolour laser from a single CdSSe bandgap-graded alloy nanoribbon (BGAN). Utilizting multiple scatters, several localized cavities are formed on the BGAN, and lasing at various colours are realized from different sections of the nanoribbon, with wavelength tuning ranging from 578 nm to 640 nm.

The CdSSe BGANs are synthesized via source-moving thermal evaporation method. The sources are CdS and CdSe powders. For a more detailed description of the process, please refer to Ref. 14. Figure 1 (a) shows the scanning electron microscopy (SEM) image of a single CdSSe BGAN we use in our experiment. Other than two visible defects and one indent, the BGAN shows smooth surface and uniform thickness as 200 nm, even though the ingredients vary continuously along the nanoribbons. It has a length of 127 μm, and an average width of 3 μm.

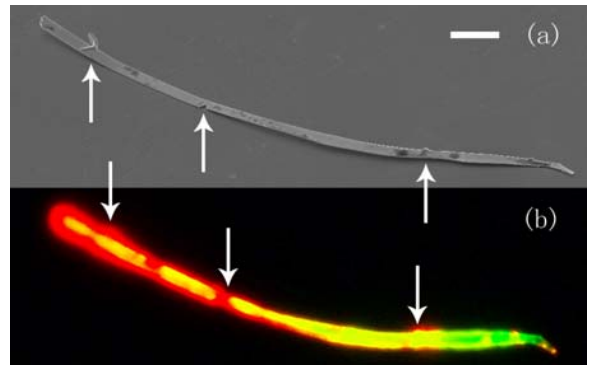

FIG. 1. (Color online) (a) SEM image of the selected bandgap-graded alloy CdSSe nanoribbon. (b) PL image of the same nanoribbon illuminated by a defocused laser beam at 532 nm. White arrows indicate the positions of defects and indent on the nanoribbons. Scale bar, 10μm.

The BGAN is first transferred from the silicon wafer grown on by a fiber taper to a glass slide. A pulsed Nd:YAG (yttrium aluminum garnet) laser with 5 ns pulse duration, 2 kHz repetition rate at 532 nm is used as excitation source. The laser is focused via a 20 X object lens (NA = 0.4), and irradiates the BGAN perpendicularly. The photoluminescence (PL) emission of the BGAN is then collected by the same object and directed to a CCD and spectrometer. A notch filter with blocking wavelength centered at 532 nm is used to block the excitation laser pulses when taking images and spectra. The details of the optical excitation-collection experimental setup can be found in Ref. 17. Figure 1 (b) shows a PL image of the BGAN, where the pump laser spot can cover the whole nanoribbon. From this image, we can find the colour of the PL varies continuously from red to green along the nanoribbon, which indicates the gradual change of the components, as well as the bandgap.

Due to the monotonous change of the bandgap along the nanoribbon, such BGAN structure usually has the property that PL generated in the middle section of the BGAN can only propagate freely to one of the end, where the bandgap is widest. However, along the bandgap-decreasing direction, due to the strong confinement of the optical mode, the guided light will be efficiently reabsorbed, and the reemission shows a red shifts.[16] In our case, red light can propagate to both ends, while the green light can only be observed at widest bandgap end. Thus one may think that, other than at long wavelength range, lasing based on axial Fabry–Pérot (FP) waveguide modes would be difficult to be realized from single BGAN structure, due to the huge reabsorption.

In our experiment, we focus the laser beam to spot of 25 μm in diameter, so that only one fifth of the nanoribbon is excited. When excitation intensity is increased above the threshold, several sharp lasing peaks emerge from broad PL background. And interestingly, depending on which section of the BGAN being excited, lasing with different colours can be realized. As shown in Fig. 2, lasing peaks centered around 578 nm, 598 nm, 621 nm, and 640 nm are observed by moving the excitation spot along the nanoribbons from wide bandgap end to the narrow bandgap end. The lasing wavelength tuning range $\Delta\lambda$ is 61 nm, which covers most of the PL spectrum range of the BGAN (plotted as black curve in Fig. 2). Considering the central wavelength of the gain is around 610 nm, we have achieved a large tunability $\Delta\lambda/\lambda^9$ of 10% from a single semiconductor nanoribbon.

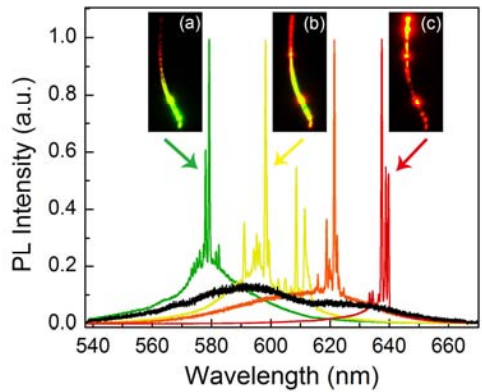

FIG. 2. (Color online) The normalized lasing spectra (shown in green, yellow, orange, red lines) collected by changing excitation spot along the nanoribbon. The black line represents the PL spectrum shown in Fig. 1 (b), for comparison. The insets are the corresponding lasing images at different wavelengths.

The dependences of laser intensity on the pumping power at different wavelength ranges are studied to ensure the lasing behavior. As shown in the inset of Fig. 3, when the laser is focused on the narrow bandgap end of the BGAN, sharp peaks around 633 nm show up and enhance dramatically with a small increasing of pump power. This clearly indicates the onset of lasing behavior, and the threshold is measured as 10 MW / cm², as shown in Fig. 3.

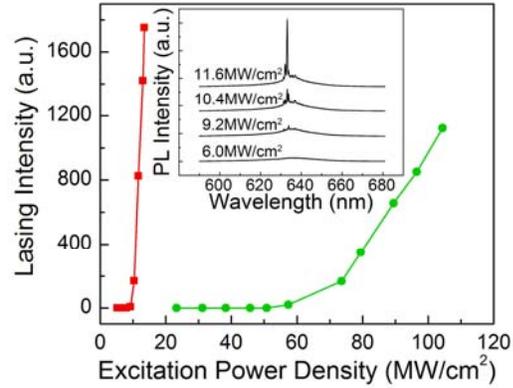

FIG. 3. (Color online) The relationships between output laser intensity and excitation power density. The red line and the green line represent the laser intensities centered at 633 nm and 579 nm, respectively. The inset shows the evolution of the lasing spectra centered at 633 nm with increasing excitation power density.

When the excitation laser is focused on the wide bandgap end, lasing is still realized around 578 nm, though the threshold is measured as 68 MW / cm² (shown in Fig. 3) and higher than that of the lasing mode at 633 nm. As we exam the corresponding lasing image, we find the laser mode does not extend to the whole nanoribbon as ordinary single nanowire FP lasers do, where the partial reflectivity at the two endfaces form the cavity. As shown in the inset (a) of Fig. 2, the bright green light is mainly localized within a section terminated by one nanoribbon end and a bright spot where a visible defect exists (indicated in Fig. 1). Beyond this section, the green light fades quickly, and PL colour changes to red due to the strong reabsorption and reemission. Thus any reflection from the other endface won't help to build up the green laser. Though defect is not favorable for a waveguide, however in this case, it acts as a "bad" mirror. In this way, the back scattering from defect provides the necessary feedback to build up the laser, and prevents the stronger absorption due to the bandgap decreasing along the nanoribbon. Nonetheless, the defect works much less efficient compare to the endface in terms of reflectivity, and this may be one of the reason that the lasing threshold at 578 nm is much higher the one that at 633 nm.

The several defects, together with two endfaces, may form many cavities on the BGAN. And this is the main reason we can tune the lasing wavelength as we move the excitation spot along the BGAN. When the excitation laser is focused on the center part of the BGAN, lasing peaks around 598 nm are observed. As shown in the inset (b) of Fig. 2, mainly three bright yellow spots exist on the BGAN, and their positions sit on one indent, one defect, and one endface. However, red colour is the leading one in the section above the indent. It is noted that for the red lasing mode peaks around 640 nm, several bright spots can be observed on the BGAN other than two endface, as shown in the inset (c) of Fig. 2. Actually, the light scattered from the lower endface is even weaker than those from other scatters. This indicates that the laser cavity at this wavelength is also different from ordinary FP nanowire lasers cavity formed by the two endfaces. Namely, multi-scattering plays the main role in the forming of laser cavity. As a result, as we can find, the lasing peaks in general show irregular spacing, more or less similar to that from random lasers.[18]

Due to the multi-scattering, lasing modes at different wavelength can be localized at different sections of the nanoribbon. Thus the reabsorption problem of the BGAN is solved, though not in a perfect way. Consequently we are able to achieve multicolour lasing simultaneously from one BGAN. By replacing the 20 X object lens with a 10 X one, we increase the excitation area to 90 μm in diameter, which covers more than half of the BGAN. Two groups of lasing peaks can be observed at the same time, and they are generated from different cavities (Fig. 4). While we move the excitation position, different lasing groups can be turned on and off. We believe that with a larger excitation area and a higher pump power, lasing from most groups within the gain range of BGAN can be realized simultaneously.

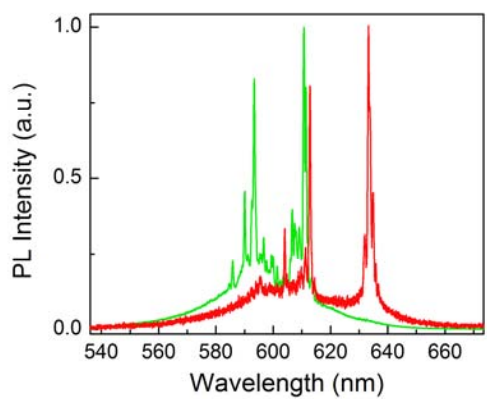

FIG. 4. (Color online) Multicolour lasing spectra from the BGAN with two excitation positions.

By employing the bandgap-graded ZnCdSSe alloy nanowire,[19] or organic photonic heterostructure nanowire,[20] together with a proper pumping source (e.g. UV pulsed laser), single BGAN white-light emitting laser with wavelength tuning range covering a wider spectrum is able to be realized. This is favorable for many integrated photonic applications such as miniature tunable laser, ultrafast nanolasers, high resolution full colour display and sensing. And introduction of an external cavity with much higher quality factor may be helpful to reach this goal.[21]

In conclusion, we have presented a multicolour tunable laser based on single CdSSe bandgap-graded alloy nanoribbon. The existence of the multiple scatters on the BGAN plays important role for realization of lasing at different wavelengths. By increasing the excitation area, multicolour laser emissions are observed simultaneously from the same nanoribbon. This work paves the way for broad BGAN applications such as miniature tunable laser, high density colour display, and sensing.

We thank Mr. Pan Wang for helpful discussion. This work is supported by the National Natural Science Foundation of China (No. 11104245), and the Fundamental Research Funds for the Central Universities.

[1]R. Yan, D. Gargas, and P. Yang, Nat. Photonics **3**, 569 (2009).
[2]J. Johnson, H. Choi, K. Knutsen, R. Schaller, P. Yang, and R. Saykall, Nat. Mater. **1**, 106 (2002).
[3]F. Qian, Y. Li, S. Gradečak, D. Wang, C. Barrelet, and C. Lieber, Nano Lett. **4**, 1975 (2004).
[4]C. Barrelet, A. Greytak, and C. Lieber, Nano lett. **4**, 1981 (2004).
[5]A. Pan, D. Liu, R. Liu, F. Wang, X. Zhu, and B. Zhou, Small **1**, 980 (2005).
[6]J. Wang, M. Gudiksen, X. Duan, Y. Cui, and C. Lieber, Science **293**, 1455 (2001).
[7]R. Yan, J. Park, Y. Choi, C. Heo, S. Yang, L. Lee and P. Yang, Nature **7**, 191 (2012).
[8]J. Zapien, Y. Liu, Y. Shan, H, Tang, C. Lee, and S. Lee, Appl. Phys. Lett. **90**, 213114 (2007).
[9]A. Pan, W. Zhou, E. Leong, R. Liu, A. Chin, B. Zou, and C. Ning, Nano Lett. **9**, 784 (2009).
[10]J. Jie, W. Zhang, I. Bello, C. Lee, and S. Lee, Nano Today **5**, 313 (2010).
[12]X. Zhuang, C. Ning, and A. Pan, Adv. Mater. **24**, 13 (2012).
[13]T. Takahashi, P. Nichols, K. Takei, A. Ford, A. Jamshidi, M. Wu, C. Ning, and A. Javey, Nanotechnology **23**, 045201 (2012).
[14]F. Gu, Z. Yang, H. Yu, J. Xu, P. Wang, L. Tong, and A. Pan, J. Am. Chem. Soc. **133**, 2037 (2011).
[15]F. Gu, H. Yu, W. Fang, and L. Tong, Appl. Phys. Lett. **99**, 181111 (2011).
[16]J. Xu, X. Zhuang, P. Guo, Q. Zhang, W. Huang, Q. Wan, W. Hu, X, Wang, X. Zhu, C. Fan, Z. Yang, L. Tong, X. Duan, and A. Pan, Nano Lett. (2012).
[17]F. Gu, H. Yu, P. Wang, Z. Yang, and L. Tong, ACS Nano **4**, 5332 (2010).
[18]H. Cao, Y. Zhao, S. Ho, E. Seelig, Q. Wang, R. Chang, Phys. Rev. Lett. **82**, 2278 (1999).
[19]Z. Yang, J. Xu, P. Wang, X. Zhuang, A. Pan, and L. Tong, Nano Lett. **11**, 5085 (2011).
[20]C. Zhang, Y. Yan, Y. Jing, Q. Shi, Y. Zhao, and J. Yao, Adv. Mater. **24**, 1703 (2012).
[21]Q. Yang, X. Jiang, X. Guo, Y. Chen, and L. Tong, Appl. Phys. Lett. **94**, 101108 (2009).

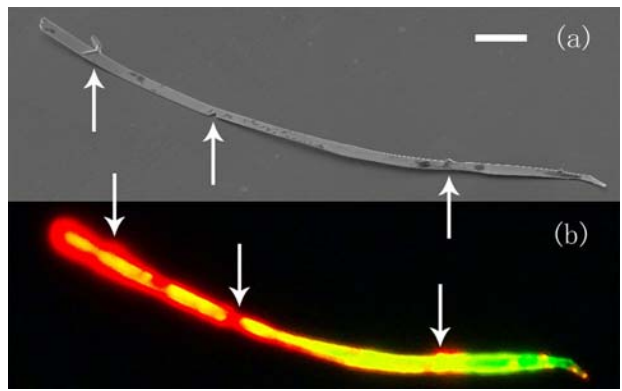

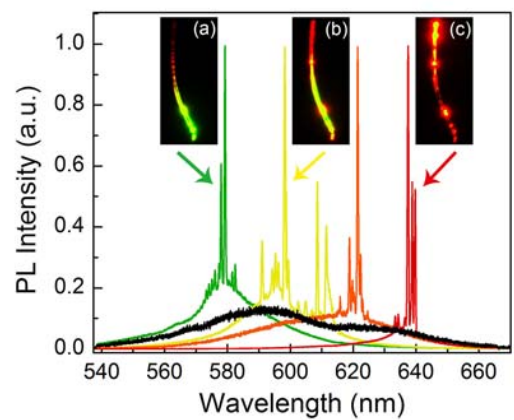

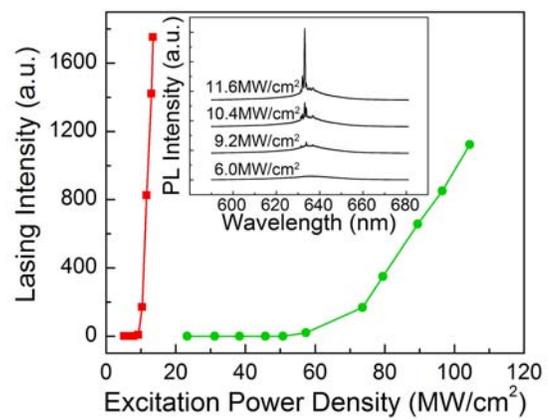

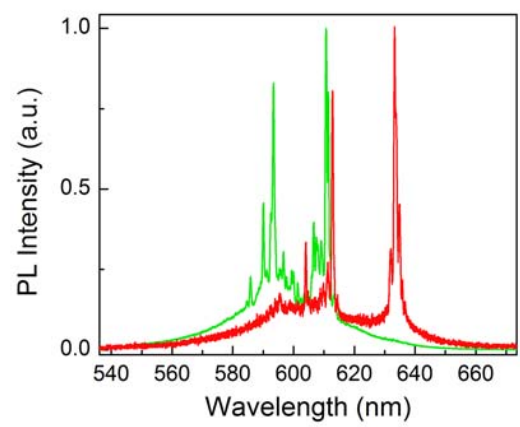